# Deep-subwavelength engineering of stealthy hyperuniformity

Jusung Park[1,2,†], Seungkyun Park[1,2,†], Kyuho Kim[2], Jeonghun Kwak[3], Sunkyu Yu[2*], and Namkyoo Park[1*]

[1]Photonic Systems Laboratory, Department of Electrical and Computer Engineering, Seoul National University, Seoul 08826, Korea

[2]Intelligent Wave Systems Laboratory, Department of Electrical and Computer Engineering, Seoul National University, Seoul 08826, Korea

[3]Department of Electrical and Computer Engineering, Inter-university Semiconductor Research Center, and SOFT Foundry Institute, Seoul National University, 1, Gwanak-ro, Gwanak-gu, Seoul 08826, Korea

[†]These authors contributed equally to this work.

E-mail address for correspondence: [*]sunkyu.yu@snu.ac.kr, [*]nkpark@snu.ac.kr

**Abstract**

Light behaviours in disordered materials have been of research interest primarily at length scales beyond or comparable to the wavelength of light, because order and disorder are often believed to be almost indistinguishable in the subwavelength regime according to effective medium theory (EMT). However, it was recently demonstrated that the breakdown of EMT occurs even at deep-subwavelength scales when interface phenomena, such as the Goos–Hänchen effect, dominate light flows. Here we develop the engineering of disordered multilayers at deep-subwavelength scales to achieve angle-selective manipulation of wave localization. To examine the disorder-dependent EMT breakdown, we classify the intermediate regime of microstructural phases between deep-subwavelength crystals and uncorrelated disorder through the concept of stealthy hyperuniformity (SHU). In this classification, we devise nontrivial order-to-disorder transitions by selectively tailoring the short-range and long-range order in SHU multilayers, achieving angle-selective control of wave localization. The result paves the way to the realization of deep-subwavelength disordered metamaterials, bridging the gap between the fields of disordered photonics and metamaterials.

# Introduction

Engineering disorder has provided multifaceted design freedom for optical devices[1]: bandgap engineering without crystalline order[2,3], imaging of biological tissues[4], transitions between ballistic and diffusive transport[5,6], deterministic control of random lasing[7,8], and disorder-induced topological transitions[9]. Compared to researches on engineering structural disorder in other domains[10,11], the uniqueness of engineering disorder in wave physics lies in utilizing wave interferences. Therefore, most studies have focused on the systems of which the characteristic lengths are beyond or comparable to the wavelength of light. At length scales below the wavelength, the phase evolution of a propagating light is negligible, which substantially degrades the impact of interferences. Therefore, a conventional way of modelling disordered materials at subwavelength scales is to introduce their effective media, rendering them almost indistinguishable from those of ordered materials.

On the other hand, the rise of subwavelength optics has transformed characteristic length scales in optics. When considerable phase evolutions occur at the interfaces through subwavelength geometry[12], the Goos–Hänchen effect[13], or non-Hermitian media[14], the necessary length scales for interferences are substantially reduced, as shown in metasurfaces[15], broken effective medium theory (EMT) in crystals[16,17], and ultrathin resonances[18]. These achievements demonstrate that a light wave can sense subwavelength microstructures by exploiting interface physics, generalizing disordered photonics into subwavelength regimes. For example, wave localization traditionally studied in the wavelength-scale[19], was observed in multilayers with deep-subwavelength characteristic lengths[20-22]. The design freedom from subwavelength disorder also stimulates advanced functionalities in scattering from metasurfaces[23-27] and light outcoupling[28].

However, most researches in this emerging field have focused on the trivial transition from order to uncorrelated disorder, such as random perturbations without any correlations[20,21,23,26,28], or conducted the black-box-type analysis: numerical assessments to achieve target functionalities[22,24,27]. To fully exploit the abundant design freedom from engineered disorder, more in-depth studies unravelling the intricate relationships between multiple wave quantities and statistical features of subwavelength materials are highly desirable. It is thus timely to address to the following question: "Can a light wave sense the correlation at deep-subwavelength scales?" Notably, recent studies on nonlocal theory for effective electromagnetic responses[29-31] underscore the need for further research, particularly in terms of engineering spatial correlations.

Here, we demonstrate that engineering disorder at deep-subwavelength scales below $\lambda/100$ enables angle-selective manipulation of wave localization, even in the simplest geometry: one-dimensional (1D) multilayers. To examine nontrivial transitions between two extremes of material phases—crystals and uncorrelated disorder—we focus on stealthy hyperuniform (SHU) multilayers and their deformations in different length scales. By analysing the EMT breakdown and angular responses of localization, we demonstrate that the spatial correlation plays a critical role even in the deep-subwavelength regime, enabling angle-selective manipulation of optical transparency through disorder engineering. The result generalizes SHU into interface physics, extends the application range of the Born approximation, and provides indispensable functionalities for high-precision sensing[32], light outcoupling[28], and radiative cooling[33].

## Results

**Deep-subwavelength engineering of multilayers**

As the simplest example of deep-subwavelength engineered disorder, we investigate a one-dimensional (1D) multilayer composed of two material phases with high and low refractive indices denoted as $n_H$ and $n_L$, respectively, as $n_H > n_L$. The multilayer is surrounded by a homogeneous material with a refractive index $n_{ext}$, and we assume the transverse electric (TE) planewave incidence from the surrounding material. To examine the impact of deep-subwavelength microstructures, we focus on a set of multilayers characterized by the identical effective refractive index $n_{EMT}$, defined as $n_{EMT} = [f_H n_H^2 + f_L n_L^2]^{1/2}$, where $f_H$ and $f_L$ denote the filling ratios of high- and low-index materials, respectively, both set at $f_H = f_L = 0.5$. It is worth mentioning that multilayers in this material set are indistinguishable with identical wave responses according to EMT[34], although it has already been disproven at two extremes—crystals[16,21] and uncorrelated disorder[20,21]—of microstructural phases, even under deep-subwavelength conditions.

In classifying materials regarding their wave properties, it is critical to employ suitable statistical parameters that properly extract the microstructural information[35], such as the perturbation strength, orders of correlations, and the clustering of constituents. In our study, we utilize the structure factor $S(k)$—the reciprocal-space representation of the two-point probability function[35,36] (Supplementary Note S1 for the calculation of structure factors). Although $S(k)$ directly determines light scattering under the first-order Born approximation, we extend our discussion beyond this approximation, as addressed later.

Despite the simplicity of multilayers considered, $S(k)$ offers various ways of exploring microstructural phases (Fig. 1a-e). At one end of the microstructural phase diagram, a 1D crystal is illustrated by the Bragg peaks at $S(k)$ (Fig. 1a). At the opposite end, there is the Poisson uncorrelated disorder, which possesses the constant $S(k)$ in the thermodynamic limit (Fig. 1e). To explore the intermediate phases between these two extremes, we employ the concept of SHU[1,29-

[31,36-38], the suppression of long-wavelength density fluctuation as $S(|k| < K) \approx 0$ for a specific positive value $K$. Notably, the SHU material exhibits the length-scale dependent degree of disorder: the crystal-like long-range (or small $|k|$) order and the Poisson-like short-range (or large $|k|$) order in terms of $S(k)$ (Fig. 1b). Therefore, we can envisage engineering disorder in terms of two distinct transitions from crystals to the Poisson disorder: degraded long-range order with broken SHU (Fig. 1c) and degraded short-range order while maintain SHU (Fig. 1d).

To investigate the interface effects on EMT breakdown, we examine oblique incidence at angle $\theta$ to engineered multilayers (Fig. 1f), which allows for angle-dependent combinations of propagating and evanescent modes within a multilayer owing to the distinct critical angle of each material phase $n_H$ and $n_L$ (Fig. 1g). Specifically, when $n_{ext} > n_{H,L}$, three distinct angles characterize the wave properties of multilayers: $\theta_H$ for the critical angle from $n_{ext}$ to $n_H$, $\theta_L$ for the critical angle from $n_{ext}$ to $n_L$, and $\theta_C$ for the critical angle from $n_{ext}$ to $n_{EMT}$. Given the relationship $\theta_L < \theta_C < \theta_H$, the angular responses of $\theta$ are classified into four regimes[20] (Fig. 1g): the fully propagating (FP) regime ($\theta < \theta_L$), the Goos-Hänchen (GH) regime ($\theta_L \leq \theta < \theta_C$), the EMT evanescent (EE) regime ($\theta_C \leq \theta < \theta_H$), and the fully evanescent (FE) regime ($\theta_H \leq \theta$). We explore wave behaviours in each angular regime in engineered disorder, extending beyond previous studies on crystals[16] and uncorrelated disorder[20,21].

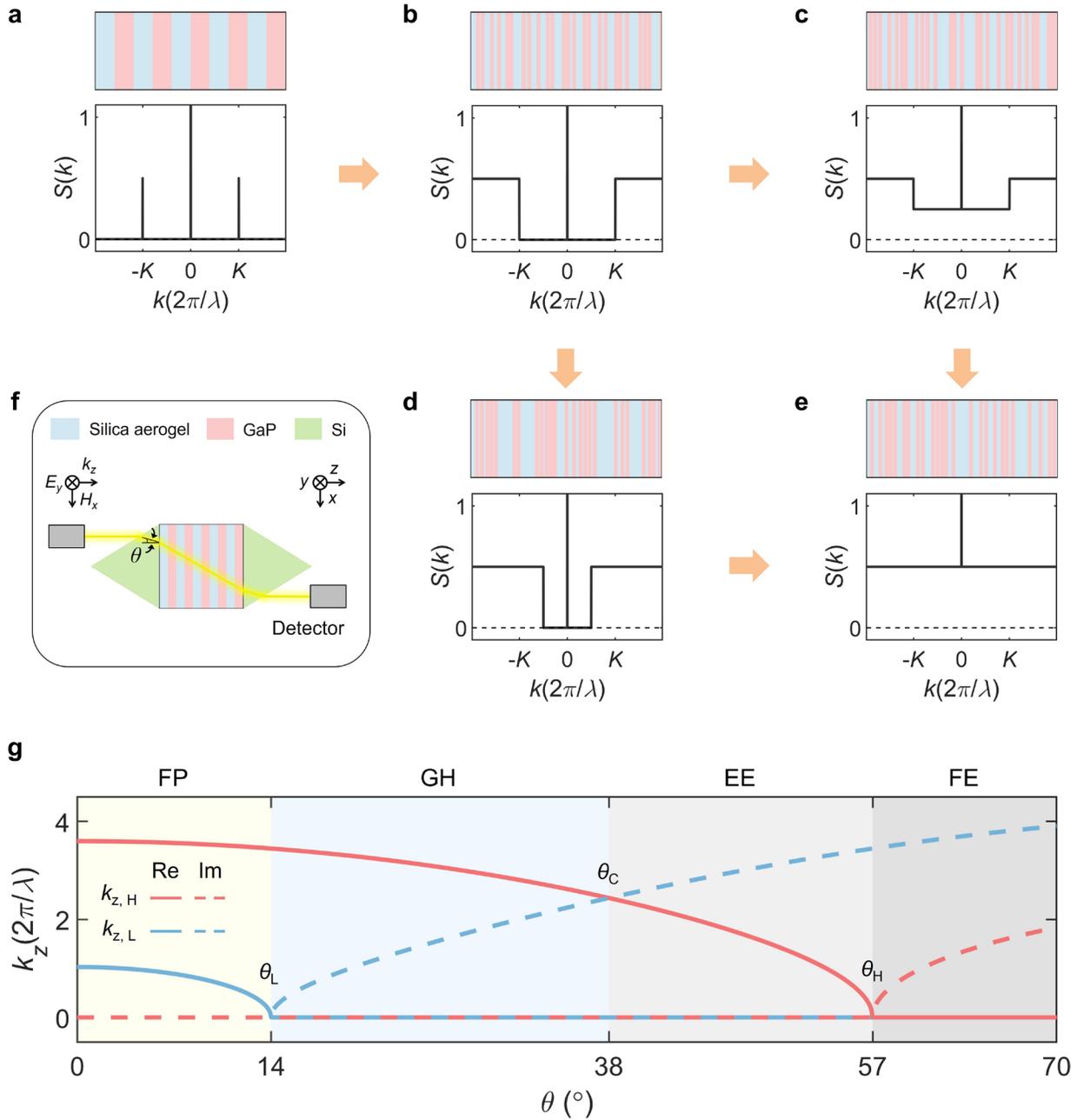

**Fig. 1. Deep-subwavelength engineered disorder. a-e**, Schematics for 1D multilayers with different degrees of disorder: crystal (**a**), SHU with larger $K$ (**b**), deformed SHU with broken long-range order (**c**), SHU with smaller $K$ (**d**), and Poisson uncorrelated disorder (**e**). The top and bottom subfigures in (**a-e**) illustrate the schematics of exemplified microstructures and their averaged $S(k)$, respectively. The orange arrows represent the order-to-disorder transitions of interest in this work. **f,** Schematic of the proposed experimental setup for measuring angular transmission. **g,** Classification of angular responses: FP ($\theta < \theta_L$), GH ($\theta_L \leq \theta < \theta_C$), EE ($\theta_C \leq \theta < \theta_H$), and FE ($\theta_H \leq$

$\theta$) regimes. In (g), we assume $n_H = 3.6$, $n_L = 1.03$, and $n_{ext} = 4.3$, which lead to $\theta_L = 14°$, $\theta_C = 38°$, and $\theta_H = 57°$. $k_{z,H}$ and $k_{z,L}$ denote the wavenumbers along the z-axis in high- and low-index materials, respectively. The details of materials for practical implementation are discussed in Methods.

**Deep-subwavelength SHU**

As the first step of exploring the intermediate regime in microstructural phases, we investigate the uniqueness of the SHU multilayers compared to crystals and uncorrelated disorder. For the incidence at a free-space wavelength of $\lambda = 500$ nm, we analyse the wave localization properties of multilayers with thickness $L$ across various material microstructures. The localization of each multilayer is quantified by the localization length $\xi$, which is defined statistically for disordered materials[39], as follows:

$$\xi = -\frac{L}{\langle \ln T(\theta) \rangle}, \quad (1)$$

where $T(\theta)$ denotes the incident-angle-dependent transmission through a realization of multilayers calculated from the scattering matrix method[40], and $\langle \ldots \rangle$ represents the ensemble average for random realizations of multilayers to examine the thermodynamic limit of disordered materials under the ergodic condition[35]. We focus on the angular range near the GH regime from $\theta_L = 14°$ to $\theta_C = 38°$, where the substantial breakdown of EMT was observed in crystals[16] and uncorrelated disorder[20,21].

Figure 2a-c shows the structure factors $S(k)$ of a crystal and the statistically designed SHU and Poisson uncorrelated disorder, which are calculated for the multilayers of the thickness from $L = 2\lambda$ to $L = 6\lambda$. While the crystal has a periodicity of 20 nm with the first-order Bragg peaks at $|k| = K = 50\pi/\lambda$ (Fig. 2a), the thickness of the high-index ($n_H$) layers in both the SHU and uncorrelated disorder is set to 2 nm. These layers are iteratively placed in a perturbative manner

inside the low-index ($n_L$) background to achieve the target $S(k)$, while maintaining $f_H = f_L = 0.5$ (see Methods and Supplementary Note S2 for the inverse design process). The design process successfully provides the target structure factors illustrated in Fig. 1a-e; the SHU satisfies $S(|k| < K) \approx 0$ and $S(|k| \geq K) \approx S_0$ (Fig. 2b), while $S(|k| > 0) \approx S_0$ in uncorrelated disorder (Fig. 2c), where $S_0$ is the constant determined by the averaged $S(k)$ in uncorrelated disorder.

Localization lengths with respect to incident angles are shown in Fig. 2d-f for each microstructural phase. Under the EMT, all the given deep-subwavelength multilayers are modelled as a homogeneous layer with the same effective index, $n_{EMT} = [(n_H^2 + n_L^2)/2]^{1/2}$. This modelling leads to angular transmissions that exhibit Fabry-Perot resonance patterns (Fig. 2d-f), where the effective wavelength is determined by the incident angles (Fig. 2g). However, although the identical $\xi(\theta,L)$ regardless of microstructures is expected in the EMT modelling, the EMT breakdown leads to the substantial decrease of $\xi(\theta,L)$ in uncorrelated disorder (Fig. 2f) compared to that of crystals (Fig. 2d).

The importance of correlation in a deep-subwavelength regime is clearly demonstrated by the localization property of SHU multilayers. Although the overall localization behaviours of the SHU look very similar to those of a crystal (Fig. 2d,e), a thorough analysis of the incident angle $\theta$ unveil the uniqueness of SHU (Fig. 2h,i). Near the boundary between the FP and GH regimes ($\theta \approx \theta_L = 14°$, Fig. 2h), the evolution of localization around varying thicknesses shows similarities in both a crystal and the SHU in terms of the widths and peak positions of $\xi(L)$. However, when the effect of the Goos-Hänchen interface phase shifts becomes more pronounced, which occurs near the boundary between the GH and EE regimes ($\theta \approx \theta_C = 38°$, Fig. 2i), localization behaviours in the SHU no longer follows those in a crystal, and instead, resemble the broadening shown in uncorrelated disorder. This result reveals that the length-scale-dependent order of the SHU— characterized by crystal-like long-range order and Poisson-like short-range order—directly affects its angular responses in the GH regime—exhibiting crystal-like localization near the FP regime and Poisson-like localization near the EE regime. Notably, because the discrepancy between the FP and EE regimes stems from the presence of evanescent modes inside multilayers (Fig. 1g), the uniqueness in the angular responses of each microstructure—a crystal, SHU, and uncorrelated disorder—originates from the distinct distributions of evanescent modes. Therefore, we can envisage more selective and deterministic engineering of optical angular functionalities near the EE regime, by independently manipulating short- and long-range order and the corresponding evanescent-mode distributions.

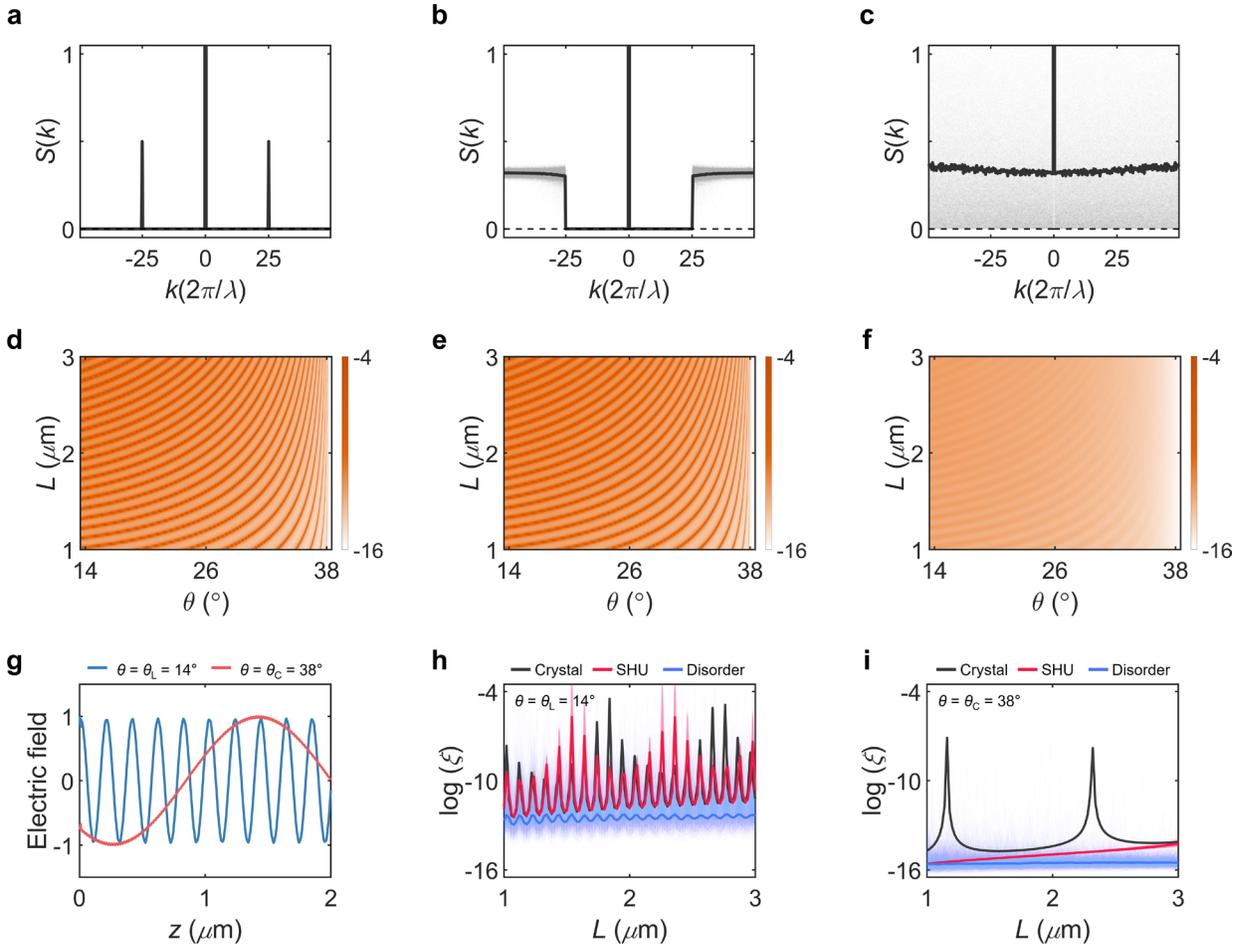

**Fig. 2. Deep-subwavelength SHU. a-c,** $S(k)$ and **d-f**, $\xi(\theta,L)$ for a crystal **(a,d)**, SHU **(b,e)**, and uncorrelated disorder (**c,f**). $L = 4\lambda$ and $K = 50\pi/\lambda$ in **a-c**. Black points and lines represent each realization and ensemble averages, respectively, in **a-c**. **g,** Normalized electric field amplitude along the *y*-axis within a crystal of $L = 4\lambda$ for $\theta = \theta_L = 14°$ and $\theta = \theta_C = 38°$. **h,i,** Localization lengths $\xi$ as a function of $L$ for $\theta = \theta_L = 14°$ (**h**) and $\theta = \theta_C = 38°$ (**i**). Transparent and solid lines denote each realization and their ensemble averages, respectively. In analyzing SHU (**b,e,h**) and uncorrelated disorder (**c,f,i**), ensembles of $10^3$ realizations are examined. All the other parameters are the same as those in Fig. 1.

**Nontrivial order-to-disorder transitions**

To employ the concept of engineered disorder[1] to impose optical functionalities on deep-subwavelength microstructures, we investigate microstructural phase transitions between the SHU and uncorrelated disorder and their impact on localization (Fig. 3). When we apply the inclusion of uncorrelated perturbations to the SHU state, the value of $S(k)$ increases across the entire range of $k$, leading to the trivial transition from SHU to uncorrelated disorder and the consequent gradual changes in wave behaviours. Instead of such a trivial transition, we devise the nontrivial transitions that are clearly differentiated by their short-range and long-range order, as well as by the maintenance of SHU. The first transition is characterized by increasing $S(|k| < K)$, which corresponds to the breakdown of long-range order (Fig. 3a→3b→3d with Fig. 3e). At the same time, the multilayers are no longer SHU because $S(|k| < K) \neq 0$. In contrast, the second transition is designed with the decrease of $K$, which degrades the short-range order of multilayers, while maintaining their SHU with $S(|k| < K) \approx 0$ (Fig. 3a→3c→3d with Fig. 3f).

Figure 3 illustrates that these two nontrivial transitions deliver distinct evolutions in maintaining Fabry-Perot resonances. At the initial SHU state (Fig. 3a), we observe distinct Fabry-Perot resonances as already depicted in Fig. 2b, while the peak positions and angular widths vary across the FP, GH, and EE regimes due to changes in effective wavelengths. Through the first-type transition with increasing $S(|k| < K)$, the breakdown of SHU leads to a highly sensitive annihilation of Fabry-Perot resonances, with the resonance peaks decreasing by an order of magnitude from a minor perturbation level of $S(|k| < K) \approx S_0/20$ for $S(|k| \geq K) \approx S_0$ (from Fig. 3a to 3b).

In contrast, the second-type transition, which involves decreasing $K$ while maintaining SHU, enables the engineering of the angular range where Fabry-Perot resonances are preserved. Notably, an abrupt change in the Fabry-Perot resonances begins to occur near the SHU state at $K = 10\pi/\lambda$ (Fig. 3c-2 and 3c-3), initiating the annihilation of the highest-order of resonance due to disrupted short-range order. The further decrease in $K$ sequentially annihilates lower-order resonances until the annihilation of the zeroth-order resonance as shown in Fig. 3d.

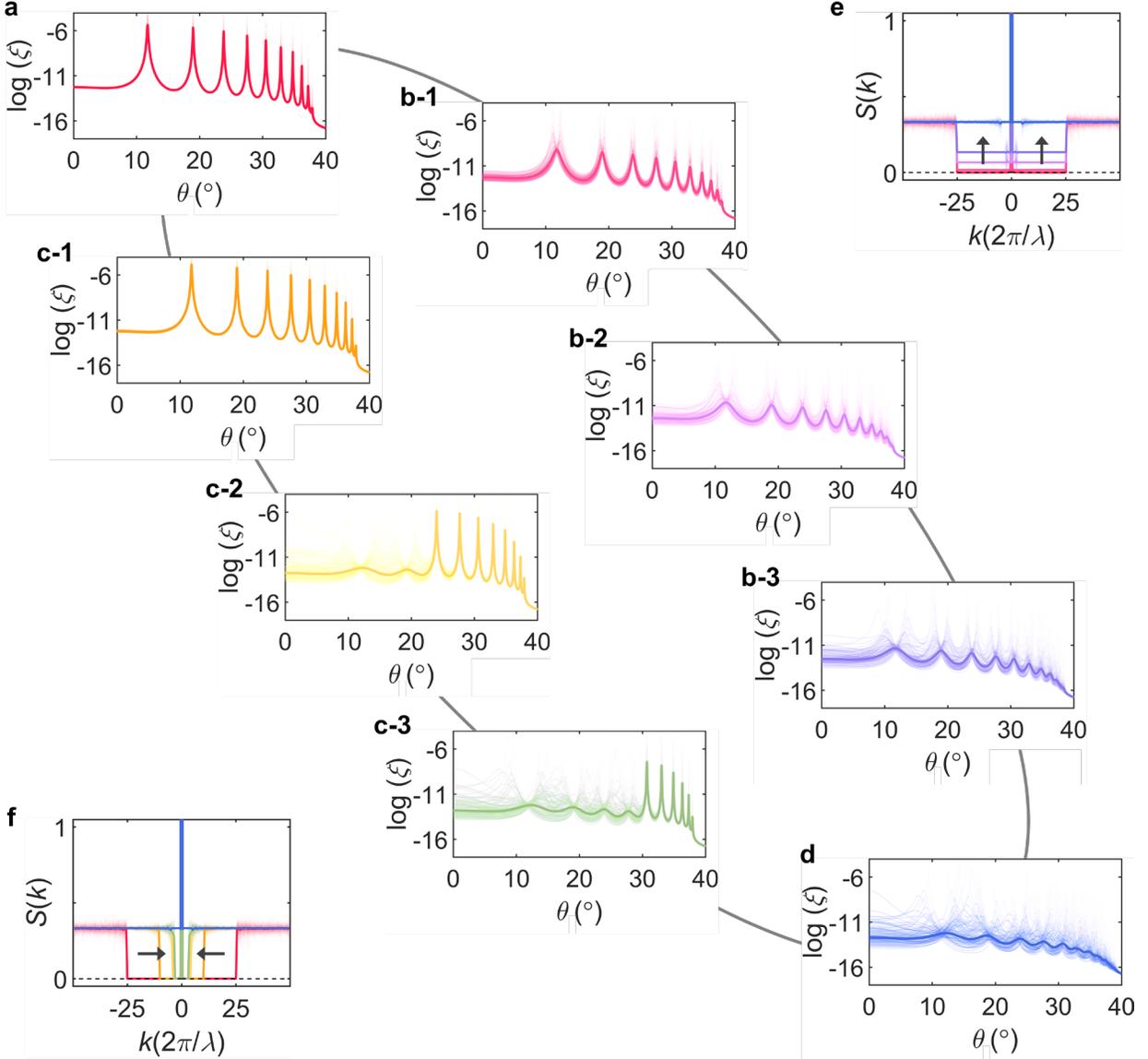

**Fig. 3. Nontrivial order-to-disorder transitions through differentiated length-scale order. a-d,** Localization lengths as functions of the incident angle $\theta$: SHU (**a**), breakdown of long-range order (**b: b-1, b-2, and b-3**), decrease of $K$ (**c: c-1, c-2, and c-3**), and uncorrelated disorder (**d**). **e,f,** $S(k)$ transitions from SHU to uncorrelated disorder for the (**b**)-transition (**e**) and (**c**)-transition (**f**). The black arrows in (**e,f**) indicate the direction of transitions for (**b-1** to **b-3**) and (**c-1** to **c-3**). In (**a-d**), points and lines represent each realization and ensemble averages, respectively. In (**e,f**), transparent and solid lines represent each realization and ensemble averages, respectively. In all cases, ensembles of $10^2$ realizations are examined. We set the entire thickness as $L = 2\lambda$. All the other parameters are the same as those in Fig. 1.

**Angle-selective engineering of localization**

Based on the results shown in Fig. 3, which demonstrate the strong connection between the density fluctuation characterized by $S(k)$ and wave localization, we further implement angle-selective engineering of localization. We employ the order-to-disorder transition at a target length scale, focusing on initial states of SHU and uncorrelated disorder. Considering the angular range of interest depicted in Fig. 3, we explore the manipulations of the SHU state near $K = 10\pi/\lambda$. Initially, the SHU multilayers exhibit Fabry-Perot resonances (Fig. 4a,b, red lines), while uncorrelated multilayers show angularly flattened strong localization (Fig. 4c,d, blue lines). To manipulate the density fluctuation at a specific length scale, we design an ensemble of disorder realizations generated by the structure factors $S(k)$ of which the value at the target $k$ increases (black line in Fig. 4a) or decreases (black line in Fig. 4c).

Through this length-scale-specific engineering of deep-subwavelength microstructures, the resonance at a specific angle can be selectively controlled over a few orders of magnitude (black lines in Fig. 4b,d). This capability to fine-tune the wave localization length can be extended to other incident angles, even allowing for precise control across multiple angles (Supplementary Note S3). Notably, when considering $L = 2\lambda$ and $n_{EMT} = 2.648$, the overall structures lie in the regime of the broken Born approximation. The results in Figs. 3 and 4 confirm that $S(k)$ serves as an excellent design tool for engineering disorder in deep-subwavelength microstructures, even when the entire system exhibits strong scattering.

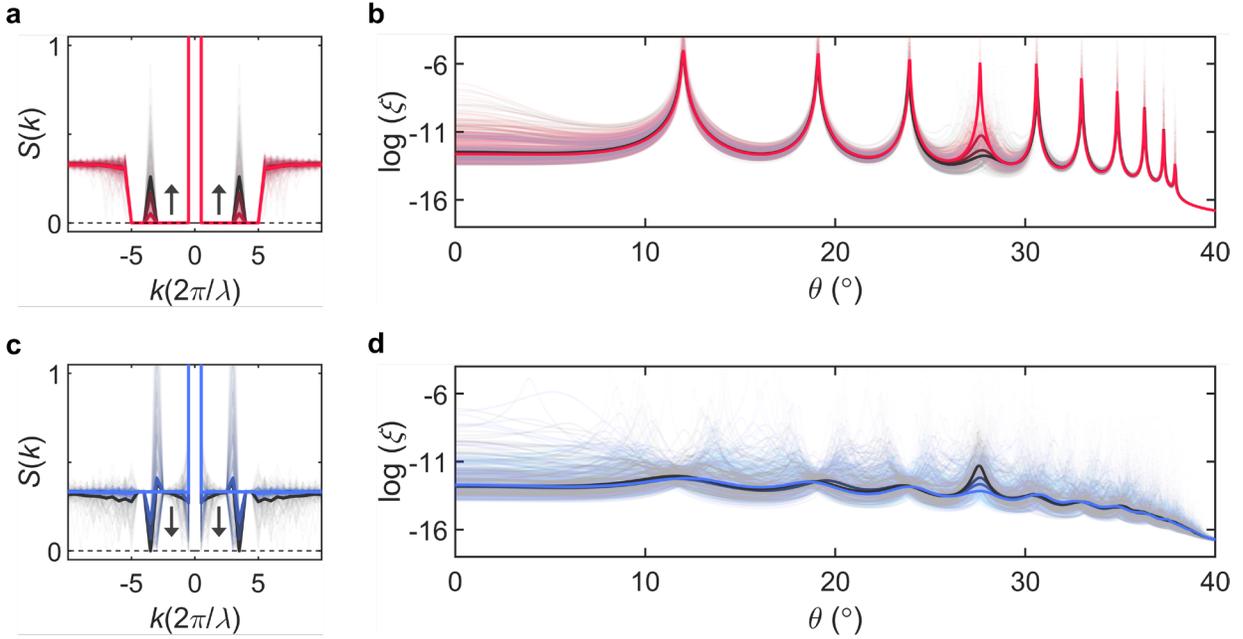

**Fig. 4. Angular-selective localization. a,b,** Selective annihilation and **c,d,** selective creation of the target Fabry-Perot resonances: the designed $S(k)$ transitions (**a,c**) and incident-angle-dependent localization lengths (**b,d**). In **a,c**, the black arrows illustrate the transition for each case. All the other parameters and plots are the same as those in Fig. 3.

**Discussion**

Achieving optical functionalities with engineered disorder in deep-subwavelength scales transforms traditional design strategies in two key aspects: designing microstructure correlations in subwavelength scales and multiple scattering regimes. First, while tailoring spatial[1,41-43] or temporal[44-46] correlations in space-time material microstructures has been widely investigated, angular optical functionalities achieved in this work reveal how interface physics can lower the length-scale boundary in microstructure correlations that affect wave phenomena substantially. Second, although our structures should be treated as multiple scattering structures, as evidenced by the multilayer thicknesses ($L = 2\lambda$ to $6\lambda$), effective index ($n_{\text{EMT}} = 2.648$), and notably low transparency (or small $\xi$), the structure factor $S(k)$ enables the deterministic design of optical functionalities at deep-subwavelength scales. Both aspects emphasize the necessity of extending the concept of disordered photonics to subwavelength optics.

Although our study exemplified a simple structure—two-phase multilayers—the underlying concept is applicable to other physical domains: spectral responses, higher-dimensional systems, time-varying photonics, and quantum optics. For instance, the proposed design directly enables spectral control of localization through the definition of structure factors, thereby inspiring the design of color selectivity at deep-subwavelength scales. We can also envision implementing two- or three-dimensional SHU systems and their deformations using deep-subwavelength elements, such as quantum dots[28]. Additionally, for GHz or THz systems, subwavelength temporal crystals and disorder can be introduced through nonlinear optics[44,47]. The recently demonstrated temporal Goos-Hänchen shift can be extended to these temporal systems[48]. Engineering in deep-subwavelength scales also inspire the connection to quantum phenomena, which will provide unique methodologies in manipulating quantum-mechanical quantities of photons[49,50].

In conclusion, we demonstrated angle-selective engineering of localization in deep-subwavelength disordered multilayers. By tailoring the profile of the structure factor $S(k)$, we achieved the inverse design of deep-subwavelength crystals, SHU, and Poisson uncorrelated disorder, as well as nontrivial transitions between these microstructural phases. We identified unique localization properties associated with each microstructural phase under the breakdown of EMT, especially near the regime of pronounced Goos-Hänchen effects at the layer interfaces. The independent manipulation of long-range and short-range order enables highly selective control of wave localization. The result paves the way for designing deep-subwavelength optical structures within the context of disorder engineering, thereby enhancing the versatility and functionalities of disordered metamaterials.

## Methods

**Deep-subwavelength multilayers.** The refractive indices of multilayers are assumed to $n_L = 1.03$ for silica aerogel layers and $n_H = 3.6$ for gallium phosphide (GaP) layers. The multilayers are surrounded by a homogenous medium composed of silicon with $n_{ext} = 4.3$. We assume the lossless condition for these materials.

**Inverse design process.** To achieve the target profile of $S(k)$ in the designed material, we employ the following iterative optimization process. First, we determine the parameters for a crystal and calculate its $S(k)$ (Figs. 1a and 2a) by following Supplementary Note S1. We also set the target structure factor $S_0(k)$ for the microstructural phase of interest (Fig. 1b-e). Second, we obtain an initial state by perturbing $10^4$ randomly sampled $n_H$ layers with replacement to enhance the convergence of the optimization method. The perturbation in preparing the initial state is uniformly random and its statistical range is set to be maximized while ensuring the hard particle condition of $n_H$ layers. We then apply $2\times10^5$ iterative optimization processes to the initial state. In each iteration, we utilize the cost function defined by the mean squared error (MSE). To enhance the convergence performance, we calculate the weighted MSE for each discretized range of the reciprocal axis $k$. First, for the reciprocal space of interest $K_I$, we divide $K_I$ into the subspaces $\kappa_j$ ($j$ = 1, 2, 3, and 4). The MSE for the $j$th subspace is defined as follows:

$$\text{MSE}_j = \frac{1}{|\kappa_j|} \int_{k \in \kappa_j} \left( \frac{S(k) - S_0(k)}{S_0(k)} \right)^2 dk, \qquad (2)$$

where $|\kappa_j|$ is the length of the subspace $\kappa_j$. Second, to reflect the distinct importance of each subspace in the structure factor, the total cost function is defined as:

$$\text{MSE}_{\text{total}} = \frac{\sum_{j=1}^{4} w_j \text{MSE}_j}{\sum_{j=1}^{4} w_j} \qquad (3)$$

where the values of the weighting factor $w_j$ are determined empirically for each microstructural phase. Finally, with the designed cost function in Eq. (3), we apply the uniformly random perturbation to $n_H$ layers with the maximum statistical range within the hard particle condition. To hinder the local minima issue, we apply the Metropolis acceptance rule[35] of which the acceptance probability is determined empirically for each microstructural phase. The proposed optimization process shows an excellent performance as shown in Figs. 2-4. An example of the optimization process is described in Supplementary Note S2.

**Data availability**

The data that support the plots and other findings of this study are available from the corresponding author upon request.

## Code availability

All code developed in this work will be made available upon request.

## Acknowledgements

We would like to thank Cheng-Wei Qiu for very helpful discussion. We acknowledge financial support from the National Research Foundation of Korea (NRF) through the Basic Research Laboratory (No. RS-2024-00397664), Young Researcher Program (No. 2021R1C1C1005031), and Midcareer Researcher Program (No. RS-2023-00274348), all funded by the Korean government. This work was supported by Creative-Pioneering Researchers Program and the BK21 FOUR program of the Education and Research Program for Future ICT Pioneers in 2024, through Seoul National University. This work was also supported by Alchemist Project grant of Korea Planning & Evaluation Institute of Industrial Technology (KEIT) funded by the Korean government (MOTIE: No. 1415185027, 20019169). We also acknowledge an administrative support from SOFT foundry institute.

## Author contributions

J.P. and S.P. contributed equally to this work. N.P. and S.Y. conceived the idea. J.P. and S.P. developed the numerical tool and performed the numerical analysis. J.P., S.P., and K.K. examined numerical analysis. J.K. discussed the practical implementation of the design. N.P. and S.Y. oversaw the project. All authors discussed the results and contributed to the final manuscript.

## Competing interests

The authors have no conflicts of interest to declare.

## Additional information

**Correspondence and requests for materials** should be addressed to S.Y. or N.P.

**Figure Legends**

**Fig. 1. Deep-subwavelength engineered disorder. a-e**, Schematics for 1D multilayers with different degrees of disorder: crystal (**a**), SHU with larger $K$ (**b**), deformed SHU with broken long-range order (**c**), SHU with smaller $K$ (**d**), and Poisson uncorrelated disorder (**e**). The top and bottom subfigures in (**a-e**) illustrate the schematics of exemplified microstructures and their averaged $S(k)$, respectively. The orange arrows represent the order-to-disorder transitions of interest in this work. **f,** Schematic of the proposed experimental setup for measuring angular transmission. **g,** Classification of angular responses: FP ($\theta < \theta_L$), GH ($\theta_L \leq \theta < \theta_C$), EE ($\theta_C \leq \theta < \theta_H$), and FE ($\theta_H \leq \theta$) regimes. In (**g**), we assume $n_H = 3.6$, $n_L = 1.03$, and $n_{ext} = 4.3$, which lead to $\theta_L = 14°$, $\theta_C = 38°$, and $\theta_H = 57°$. $k_{z,H}$ and $k_{z,L}$ denote the wavenumbers along the $z$-axis in high- and low-index materials, respectively. The details of materials for practical implementation are discussed in Methods.

**Fig. 2. Deep-subwavelength SHU. a-c,** $S(k)$ and **d-f**, $\xi(\theta,L)$ for a crystal (**a,d**), SHU (**b,e**), and uncorrelated disorder (**c,f**). $L = 4\lambda$ and $K = 50\pi/\lambda$ in **a-c**. Black points and lines represent each realization and ensemble averages, respectively, in **a-c**. **g,** Normalized electric field amplitude along the $y$-axis within a crystal of $L = 4\lambda$ for $\theta = \theta_L = 14°$ and $\theta = \theta_C = 38°$. **h,i,** Localization lengths $\xi$ as a function of $L$ for $\theta = \theta_L = 14°$ (**h**) and $\theta = \theta_C = 38°$ (**i**). Transparent and solid lines denote each realization and their ensemble averages, respectively. In analyzing SHU (**b,e,h**) and uncorrelated disorder (**c,f,i**), ensembles of $10^3$ realizations are examined. All the other parameters are the same as those in Fig. 1.

**Fig. 3. Nontrivial order-to-disorder transitions through differentiated length-scale order. a-d,** Localization lengths as functions of the incident angle $\theta$: SHU (**a**), breakdown of long-range order (**b: b-1, b-2, and b-3**), decrease of $K$ (**c: c-1, c-2, and c-3**), and uncorrelated disorder (**d**). **e,f,** $S(k)$ transitions from SHU to uncorrelated disorder for the (**b**)-transition (**e**) and (**c**)-transition (**f**). The black arrows in (**e,f**) indicate the direction of transitions for (**b-1** to **b-3**) and (**c-1** to **c-3**). In (**a-d**), points and lines represent each realization and ensemble averages, respectively. In (**e,f**), transparent and solid lines represent each realization and ensemble averages, respectively. In all cases, ensembles of $10^2$ realizations are examined. We set the entire thickness as $L = 2\lambda$. All the other parameters are the same as those in Fig. 1.

**Fig. 4. Angular-selective localization. a,b,** Selective annihilation and **c,d,** selective creation of the target Fabry-Perot resonances: the designed $S(k)$ transitions (**a,c**) and incident-angle-dependent localization lengths (**b,d**). In **a,c**, the black arrows illustrate the transition for each case. All the other parameters and plots are the same as those in Fig. 3.

on strong interference effects in highly absorbing media. *Nat. Mater.* **12**, 20-24 (2013).

19. Segev, M., Silberberg, Y. & Christodoulides, D. N. Anderson localization of light. *Nat. Photon.* **7**, 197-204 (2013).

20. Herzig Sheinfux, H., Kaminer, I., Genack, A. Z. & Segev, M. Interplay between evanescence and disorder in deep subwavelength photonic structures. *Nat. Commun.* **7**, 12927 (2016).

21. Sheinfux, H. H., Lumer, Y., Ankonina, G., Genack, A. Z., Bartal, G. & Segev, M. Observation of Anderson localization in disordered nanophotonic structures. *Science* **356**, 953-956 (2017).

22. Oh, S., Kim, J., Piao, X., Kim, S., Kim, K., Yu, S. & Park, N. Control of localization and optical properties with deep-subwavelength engineered disorder. *Opt. Express* **30**, 28301-28311 (2022).

23. Yu, Z., Li, H., Zhao, W., Huang, P.-S., Lin, Y.-T., Yao, J., Li, W., Zhao, Q., Wu, P. C. & Li, B. High-security learning-based optical encryption assisted by disordered metasurface. *Nat. Commun.* **15**, 2607 (2024).

24. Xu, M., He, Q., Pu, M., Zhang, F., Li, L., Sang, D., Guo, Y., Zhang, R., Li, X. & Ma, X. Emerging long-range order from a freeform disordered metasurface. *Adv. Mater.* **34**, 2108709 (2022).

25. Zhang, H., Cheng, Q., Chu, H., Christogeorgos, O., Wu, W. & Hao, Y. Hyperuniform disordered distribution metasurface for scattering reduction. *Appl. Phys. Lett.* **118**, 101601 (2021).

26. Maguid, E., Yannai, M., Faerman, A., Yulevich, I., Kleiner, V. & Hasman, E. Disorder-induced optical transition from spin Hall to random Rashba effect. *Science* **358**, 1411-1415
24

# Supplementary Information for "Deep-subwavelength engineering of stealthy hyperuniformity"

Jusung Park[1,2,†], Seungkyun Park[1,2,†], Kyuho Kim[2], Jeonghun Kwak[3], Sunkyu Yu[2*], and Namkyoo Park[1*]

[1]Photonic Systems Laboratory, Department of Electrical and Computer Engineering, Seoul National University, Seoul 08826, Korea

[2]Intelligent Wave Systems Laboratory, Department of Electrical and Computer Engineering, Seoul National University, Seoul 08826, Korea

[3]Department of Electrical and Computer Engineering, Inter-university Semiconductor Research Center, and SOFT Foundry Institute, Seoul National University, 1, Gwanak-ro, Gwanak-gu, Seoul 08826, Korea

[†]These authors contributed equally to this work.

E-mail address for correspondence: [*]sunkyu.yu@snu.ac.kr, [*]nkpark@snu.ac.kr

**Note S1. Structure factor calculation**

**Note S2. Iterative optimization process**

**Note S3. Localization length with multiple angular selectivity**

**Note S1. Structure factor calculation**

The structure factor $S(k)$ in a one-dimensional (1D) inhomogeneous material of length $L$ for the refractive index profile $n(z)$ is defined as follows:

$$S(k) = \frac{1}{L}\left|\int_0^L n^2(z)e^{-ikz}dz\right|^2. \qquad (S1)$$

To analyse the two-phase material composed of the binary-value refractive indices $n_H$ and $n_L$, we model the material as a series of the $n_H$ layer embedded in the background material of $n_L$ (Fig. S1). The Fourier transform of $n^2(z)$ then becomes:

$$\begin{aligned}
F\{n^2(z)\} &= \int_{-\infty}^{\infty} n^2(z)e^{-ikz}dz \\
&= (n_H^2 - n_L^2)F\left\{\sum_{j=1}^{l}\delta(z-z_{H,j}) * \mathrm{rect}\left(\frac{z}{d_{H,j}}\right)\right\} + F\{n_L^2\} \\
&= (n_H^2 - n_L^2)\left(\sum_{j=1}^{l}F\{\delta(z-z_{H,j})\}F\left\{\mathrm{rect}\left(\frac{z}{d_{H,j}}\right)\right\}\right) + F\{n_L^2\} \\
&= (n_H^2 - n_L^2)\left(\sum_{j=1}^{l}e^{-ikz_{H,j}}d_{H,j}\mathrm{sinc}\frac{kd_{H,j}}{2\pi}\right) + n_L^2\delta(k=0),
\end{aligned} \qquad (S2)$$

where $F$ denotes the Fourier transform, $\mathrm{rect}(z/d_{H,j})$ is rectangular function of the $j$th $n_H$ layer with the width $d_{H,j}$, the symbol '*' denotes the convolution, $\delta$ represents the Dirac delta function, $z_{H,j}$ is the center of the $j$th rectangular function, and $l$ is the number of the $n_H$ layers. Considering the thermodynamic limit of $L \to \infty$, the structure factor for the finite-length material is calculated as follows:

$$S(k) = \frac{1}{L}\left|(n_H^2 - n_L^2)\left(\sum_{j=1}^{l}e^{-ik \cdot z_{H,j}}d_{H,l}\mathrm{sinc}\frac{kd_{H,l}}{2\pi}\right) + n_L^2\delta(k=0)\right|^2. \qquad (S3)$$

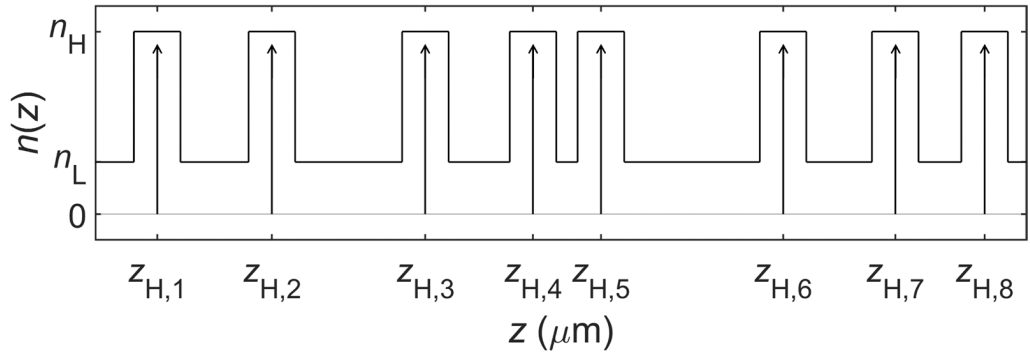

**Fig. S1. Refractive index profile for two-phase materials.** The optical potential is described by the refractive index profile $n(z)$, which is composed of the rectangular functions that describe the $n_H$ layers.

**Note S2. Iterative optimization process**

Figure S2 describes an example of the optimization process for the structure factor of 1D SHU in Figure 2b in the main text. While Fig. 2a illustrates the decrease of the cost function defined by the MSE (see Methods), Fig. 2b shows an excellent agreement between the target structure factor and the obtained structure factors. Notably, the design process can also be considered including the randomly perturbed Fourier transforms, as described in Supplementary Note S1.

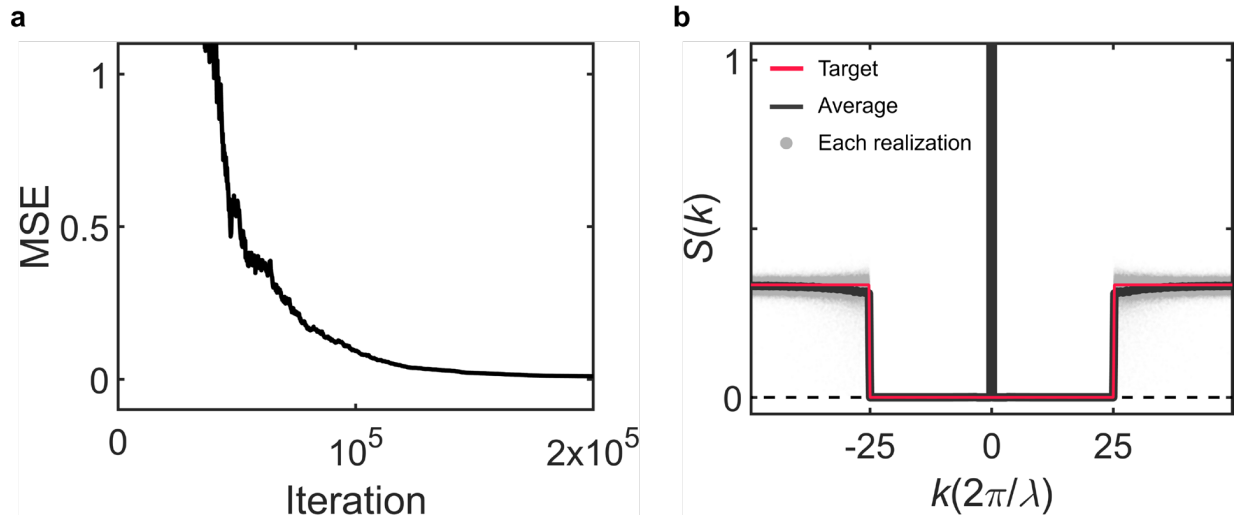

**Fig. S2. $S(k)$ design example. a,** The evolution of the MSE during the iterative optimization processes. **b,** The target (red line) and the obtained (black points for each realization and black line for the average) structure factor. $L = 4\lambda$. An ensemble of $10^3$ realizations is examined.

**Note S3. Multiple angle selectivity**

By controlling $S(k)$, we can manipulate the localization length for multiple incident angles (Fig. S3). Starting from the SHU peaks at $K = 11\pi/\lambda$, we adjust the $S(k)$ at $|k| = 9\pi/\lambda$ and $7\pi/\lambda$ (Fig. S3a,c). The corresponding incident angles of the controlled localization length are $\theta = 24°$ and $31°$ (Fig. S3b,d). The increase and decrease of the target length-scale density fluctuations lead to the suppression (Fig. S3b) and enhancement (Fig. S3d) of the localization length, respectively, demonstrating the ability to realize multiple-angle selectivity in controlling wave localization.

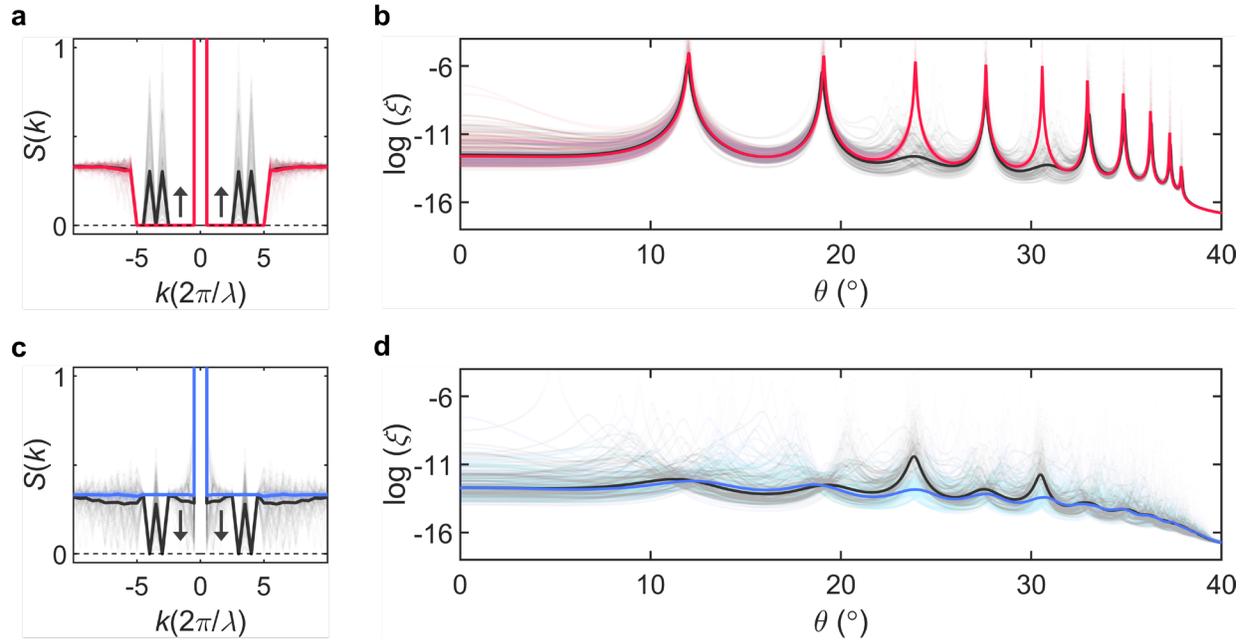

**Fig. S3. Multi-angle selective localization. a,b,** Selective annihilation and **c,d,** selective creation of the target Fabry-Perot resonances: the designed $S(k)$ transitions (**a,c**) and incident-angle-dependent localization lengths (**b,d**). In **a,c**, the black arrows illustrate the transition for each case. All the other parameters and plots are the same as those in Fig. 3 in the main text.